\documentclass[aps,preprint,groupedaddress]{revtex4-1}
\usepackage{float}
\usepackage{multirow}
\usepackage{amsmath}
\usepackage{graphicx}
\usepackage{stackrel}
\usepackage{natbib}
\usepackage[unicode=true]{hyperref}

\begin{document}
	\title{Speeding up of Binary Mergers due to `Apparent' Gravitational
		Wave Emissions}
	
	\author{Shibaji Banerjee}
	\email{shiva@sxccal.edu}
	\affiliation{Department of Physics, St. Xavier's College, 30, Mother Teresa Sarani, Kolkata-700016, India.}
	
	\author{Ashadul Halder}
	\email{ashadul.halder@gmail.com}
	\affiliation{Department of Physics, St. Xavier's College, 30, Mother Teresa Sarani, Kolkata-700016, India.}
	
	\author{Sanjay K. Ghosh}
	\email{sanjay@jcbose.ac.in}
	\affiliation{Department of Physics, Unified Academic Campus, Bose Institute, EN-80, Sector V, Bidhannagar, Kolkata-700091, India.}
	
	\author{Sibaji Raha}
	\email{sibaji.raha@jcbose.ac.in}
	\affiliation{Department of Physics, Unified Academic Campus, Bose Institute, EN-80, Sector V, Bidhannagar, Kolkata-700091, India.}
	
	\author{Debasish Majumdar}
	\email{debasish.majumdar@saha.ac.in}
	\affiliation{Astroparticle Physics and Cosmology Division, Saha Institute of Nuclear Physics, HBNI\\1/AF Bidhannagar, Kolkata-700064, India.}
	
	\begin{abstract}
		Gravitational waves from binary black hole pairs \cite{binary_bh_old} have emerged as an important observational tool in current times. The energy of the BH - BH binary pair is radiated in the form of gravitational waves and to compensate for that energy, kinetic energy of the system decreases gradually. Consequently the mutual separation of the objects decreases with time and tends to merge. The whole process may require a very long time comparable or longer than the age of the universe, specially in the case of low mass mergers. We have examined the case in which a massive object compared to the individual masses comprising the binary pair is present nearby such a system. We have found that in this case the merging process takes place much rapidly than that of the conventional BH-BH merging process \cite{wheeler,schutz,h_and_L,tapir_caltech}. Scenarios with both an Intermediate Mass Black Hole (IBMH) ($10^{5}\:M_{\odot}$) as well as a Super Massive Black Hole (SMBH) have been studied and the latter has been found to provide a much higher overall rate for the BH-BH merger process.
	\end{abstract}

	\pacs{}
	\maketitle

	\section{Introduction}
	Almost hundred years after Einstein discovered the existence of gravitational waves mathematically in the context of the General Theory of Relativity they are increasingly turning out to be the ubiquitous messengers bringing ``light'' from the dark corners of the universe.  According to the basic premises of relativity, gravitational waves are nothing but the disturbance in space-time, propagating in space in the form of ripples with the velocity of light. As those ripples pass, they deform the adjacent space in periodic fashion. These sorts of deformations were discovered by the LIGO and VIRGO collaboration (GW150914) \cite{binary_bh_old} in 2014, which had the first ever direct detection of gravitational waves, beginning a new chapter in modern physics. 
	
	These newfound waves has several similarities to the electromagnetic waves, though they contain some special characteristics which enable the gravitational waves capable to investigate numerous unobserved astrophysical events. The gravitational waves remain unscathed \cite{unscathed,unscathed_1} during its propagation and being quadrupolar in nature, the intensity falls off only inversely with the distance. As a result, it seems to be the only medium to observe the highly red shifted and compact astrophysical events. Also the gravitational waves play an important role in understanding primordial cosmology \cite{primordial,pre_inflation}.
	
	According to the notion of relativity, everything in the universe are embedded in that four dimensional space time, which is partially deformed by gravity waves during its propagation. So, it may retains significant impacts in their regular characteristics/ motions. As a consequences, in searching for highly red shifted and low luminous events, the insufficiency of the direct optical observations can be repealed by the gravitational wave astronomy. An unmoving Intermediate Mass Black Hole (IMBH) is the one of such candidates, which is almost impossible to locate in direct optical observation for their low luminosity and lightweights. Consequently, the most relevant way to discover their existence is to look for the effect of them on neighboring events.
	
	Recent observations on low-luminosity active galactic nuclei (AGN) seems to provide strong evidence in favor of the existence of Intermediate Mass Black Hole (IMBH) \cite{m15imbh} of masses $\sim10^{4}\:M_{\odot}$. N-body simulation of Baumgardt et al. \cite{Nbody} and dynamical modeling of Gerssen et al. \cite{dmodel} suggest the existence of one or more central IMBHs in non core-collapsed clusters. Any binary star system located near the center of such clusters and galaxies may be affected by those IMBHs. In this case, in spite of being too lightweight to radiate gravitational wave with significant intensity, the binary stars may collide due to the presence of the nearby IMBHs or SMBHs. The BH binaries, along with a massive neighbor comprise a three body system as in HD 181068 \cite{HD_181068}, HD 188753 \cite{HD_188753} and Alpha Centauri \cite{Alpha_Centauri}. In the present work, we explore the possibility that the collision time for a binary is reduced due to the influence of an IMBH. This is essentially due to the fact that the effective three body system loses power through a gravitational wave mechanism at enhanced rate relative to the binary system.
	
	The paper is organized as follows. In section \ref{twobody} we give an account of a two body system and their eventual merger. In section \ref{threebody} we extend our formalism for a three body system, where a binary system rotates around a massive third body. Section \ref{gwmergerke} describes the formalism for the gravitational waves from such a three body system and the eventual loses of kinetic energy of the same system. In section \ref{csandrs} we furnish the calculational procedures and the results. Finally in section \ref{conclu} some concluding remarks are given.
	
	\section{\label{twobody}Two body system and conventional BH-BH merging}
	
	The two-body problem encompasses the motion of two bodies, where the
	only interaction considered is the gravitational force due to their
	masses and any other interaction from any other force is neglected.
	Binary stars, motion of planet around stars, motion of moons around
	planet etc. are some good examples of such a system. 
	
\begin{figure}
		\centering{}
		\caption{\label{fig:unigrav}Law of universal gravitation}
		\includegraphics[scale=0.5]{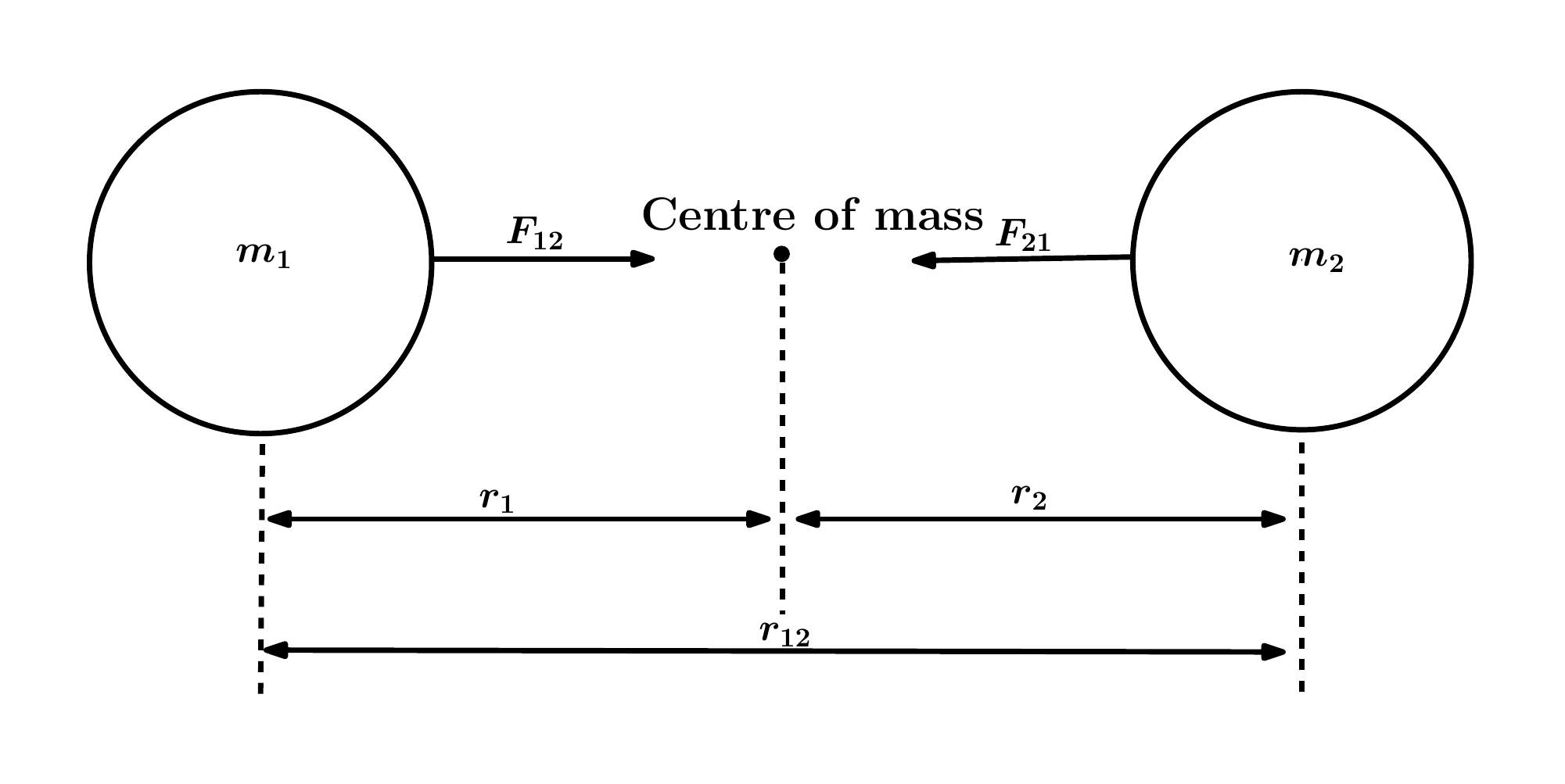}
	\end{figure}

	The astrodynamical two body problem (Fig. \ref{fig:unigrav}) deals
	with the motion of two astronomical bodies, exemplified by motion
	of planets and moons around the sun or a planet. The dynamics of a binary
	star system fall within this class of celestial motion in which the
	participants are governed solely by their gravitational interactions
	which depend only on their masses both in a Newtonian or a General
	Relativistic scenarios. 
	
	Let, $\boldsymbol{r}_{1}(t)$, $\boldsymbol{r}_{2}(t)$ be the position vectors of the two bodies with respect to their common center of mass having masses $m_{1}$ and $m_{2}$ respectively at a given time $t$ . According to the Newton's law of universal gravitation the gravitational force $\boldsymbol{F}_{12}$, acting on the first body due to the second body is given by, 
	
	\[
	\boldsymbol{F}_{12}=\frac{Gm_{1}m_{2}}{\mid r_{12}\mid^{3}}\boldsymbol{r_{12}},
	\]
	
	\noindent where $G$ is the universal gravitational constant and  $\boldsymbol{r}_{12}=\boldsymbol{r}_{2}-\boldsymbol{r}_{1}$.
	According to the Newton's second law, an object revolving around the center of mass of the system in a circular orbit with constant angular velocity $\omega$ is subjected to the centripetal force $\boldsymbol{F}_{12}$ with, 
	
	\begin{equation}
	\boldsymbol{F}_{12}(\boldsymbol{r}_{1},\boldsymbol{r}_{2})=m_{1}\ddot{\boldsymbol{r}}_{1}=-m_{1}\omega^{2}\boldsymbol{r}_{1},\label{Newton Law 12}
	\end{equation}
	
	\begin{equation}
	\boldsymbol{F}_{21}(\boldsymbol{r}_{1},\boldsymbol{r}_{2})=m_{2}\ddot{\boldsymbol{r}}_{2}=-m_{2}\omega^{2}\boldsymbol{r}_{2}.\label{Newton Law 21}
	\end{equation}
	
	Addition of the force Eqs. (\ref{Newton Law 12}) and (\ref{Newton Law 21})
	yields
	\begin{equation}
	m_{1}\ddot{\boldsymbol{r}}_{1}+m_{2}\ddot{\boldsymbol{r}}_{2}=(m_{1}+m_{2})\ddot{\boldsymbol{R}}=\boldsymbol{F}_{12}+\boldsymbol{F}_{21}=0,\label{Newton total force}
	\end{equation}
	
	\noindent where $\boldsymbol{R}$ locates the center of mass of the system and $\omega$ is the angular velocity given by 
	\begin{equation}
	\omega=\sqrt{\frac{G(m_{1}+m_{2})}{\mid\boldsymbol{x}_{12}\mid^{3}}}.\label{omega_old}
	\end{equation}
	
	From Eq. (\ref{Newton total force}) we have $\ddot{\boldsymbol{R}}=0$,
	which implies that the center of mass of the system moves with uniform velocity (or is at rest). Consequently, both the bodies revolve around the center of mass of the system with same angular velocity while kinetic energy of the binary system remains unchanged. 
	
	\begin{figure}
		\centering{}
		\caption{Orbit of two body system. In the present work $m_{1}=m_{2}=m$ is chosen.}
		\includegraphics[scale=0.6]{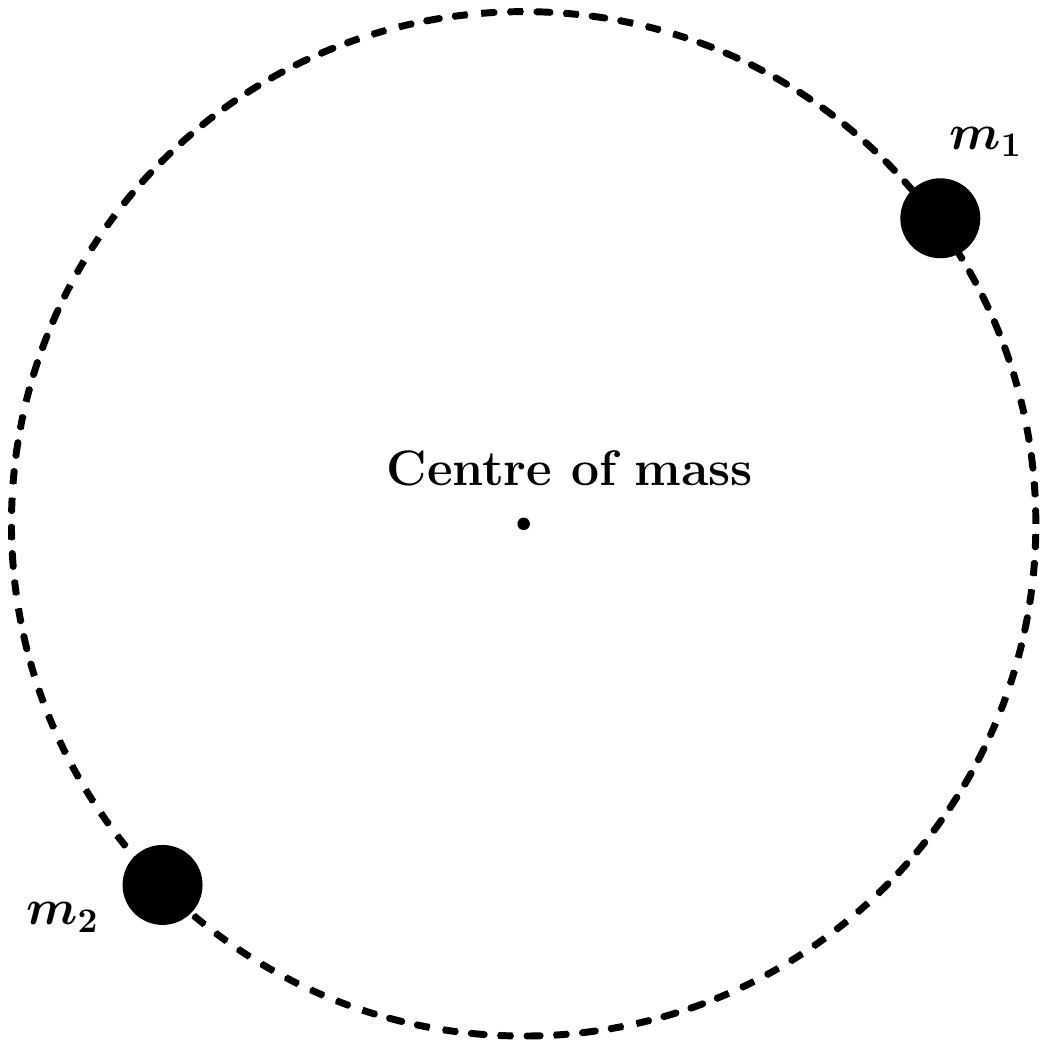}
	\end{figure}
	
	However, when the kinetic energy changes with time as a result of any internal or external interaction, orbits of both the particles and velocities assume new equilibrium values. Black Hole - Black Hole (BH-BH) merger in the context of gravitational waves is a well known example of it, in which two massive astronomical objects form a binary system and orbits each other. According to the General Theory of Relativity \cite{wheeler} any massive object curves the space-time around it. So, the change in the orbital motion of such a massive object gives rise to a ripple in space-time fabric, which is known as gravitational wave (GW). GW carries energy and propagates in space with the velocity same as that of the light like electromagnetic (EM) wave.
	
	An equal mass binary BH system consists of two black holes of equal masses of magnitude $m$ and orbital radius $r$, would trace a path along equivalent orbits around their center of mass with angular velocity $0.5\times(Gmr)^{1/2}$. As a result, the system would radiate gravitational waves having average luminosity $\mathcal{L}_{\text{GW}}=\dfrac{64}{5}\dfrac{G^{4}}{c^{5}}\dfrac{m^{5}}{(2r)^{2}}$  \cite{wheeler,schutz,h_and_L,tapir_caltech} as the outcome of the change in quadrupole moment of its own with time. In the adiabatic regime, both BHs of the system would adjust their own velocity as well as orbital radius in order to make up for the energy loss due to gravitational wave emission in such a way that the relation \cite{wheeler,schutz,h_and_L,tapir_caltech}
	
	\begin{equation}
	\mathcal{L}_{\text{GW}}+\frac{d}{dt}E_{k}=0\label{eq:equilibrium}
	\end{equation}
	
	\noindent is satisfied. In the above, $\mathcal{L}_{\text{GW}}$ is the gravitational wave luminosity and $E_{k}$ is the total kinetic energy of the system. The above equation (Eq. \ref{eq:equilibrium}) implies that as the kinetic energy of the binary decreases with time along with the orbital velocities, the mutual separation between them reduces to the minimal possible value, eventually producing a new black hole as a result of their merger. However, in the process, the angular velocities of the objects increase (Eq. \ref{omega_old}). Generally the collapse of low mass stars and exo-planets takes a very long duration of time. So only BH-BH merger processes are observable for this kind of phenomenon.
	
	\section{\label{threebody}An Effective Three body system}
	
\begin{figure}[H]
		\centering{}
		\caption{\label{fig:Triple-star-system}A triple star system consisting of a binary system and a massive star. The mass of the massive star is much larger than those of the stars of the binary. See text for details.}
		\includegraphics[scale=0.6]{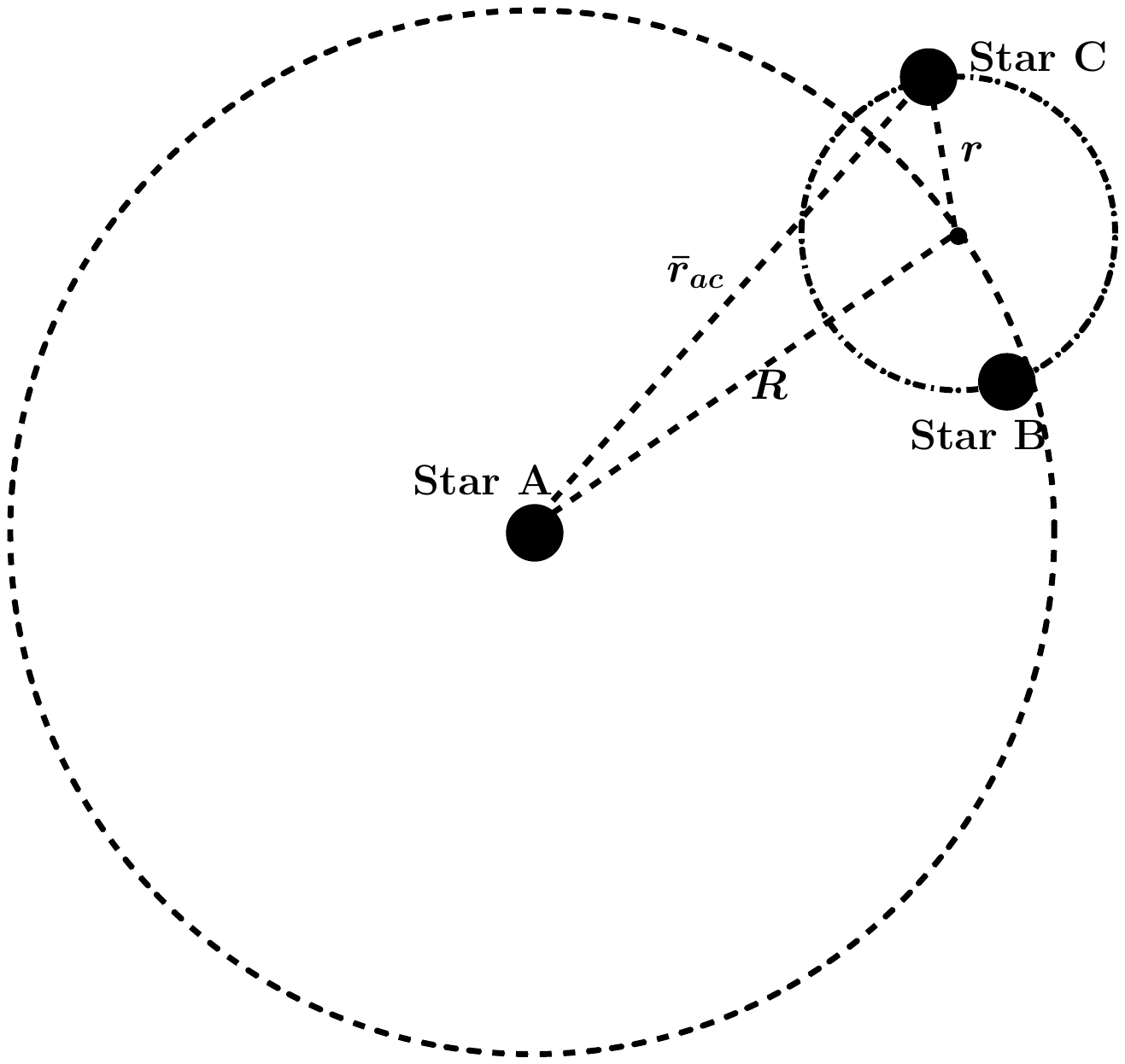}
	\end{figure}

	A three body problem \ref{fig:Triple-star-system} is concerned with the motion of three objects (such as a triple star system) under the influence of their mutual gravitational force. However, unlike the gravitational 2-body problem, no general solution is available for a three body system. We have therefore assumed a simplified and analytically tractable model of a triple-star system in this work by considering that the system consists of a primary massive star or a black hole A having mass $M$ and a binary system, consisting of stars B and C having masses $m_{\text{B}}$ and $m_{\text{C}}$ respectively, while the latter (the binary system) is in orbital motion of radius $R$ around the primary star A (Fig. \ref{fig:Triple-star-system}). The orbital radii of the stars A, B and C are $r_{a}$, $r_{b}$ and $r_{c}$ respectively. Star B and star C orbit around their common center of mass with angular velocity $\omega$ and binary system B-C orbits star A with angular velocity $\omega_{1}$. For the sake of simplicity we also assume that $m_{\text{B}}=m_{\text{C}}=m$, $r_{b}=r_{c}=r$ and $M\gg m$.
	
	The above assumptions are equivalent to assuming that $\omega\gg\omega_{1}$ (Eq. \ref{omega_old}) and that the orbital radius $r_{a}$ of star A can be ignored relative to the other orbital scales, in turn renders that the quadrupole moment of star A and the corresponding gravitational wave luminosity are negligible. But the distance between star A and star C changes with time resulting in a fluctuating gravitational quadrupole moment. Therefore gravitational waves can be observed from star B or star C due to relative motion of star A with respect to the binary system. Position vector of the CM (center of mass) of binary system observed from star A at time $t$ is given by, 
	
	\begin{equation}
	\vec{r}_{\text{CM}}=-R\sin\omega_{1}t\:\hat{i}-R\cos\omega_{1}t\:\hat{j},\label{eq:position ab}
	\end{equation}
	
	\noindent where $R$ is the mutual separation between the center of masses of star A and the binary system. Similarly, position vector of star C observed from the star B at time $t$ is given by
	
	\begin{equation}
	\vec{r}_{bc}=\vec{r_c}-\vec{r_b}=-r\sin\omega t\:\hat{i}-r\cos\omega t\:\hat{j}.\label{eq:position bc}
	\end{equation}
	
\begin{figure}[H]
		\centering{}
		\caption{\label{fig:Aperant-motion}Apparent motion of star A observed from star C}
		\includegraphics[scale=0.6]{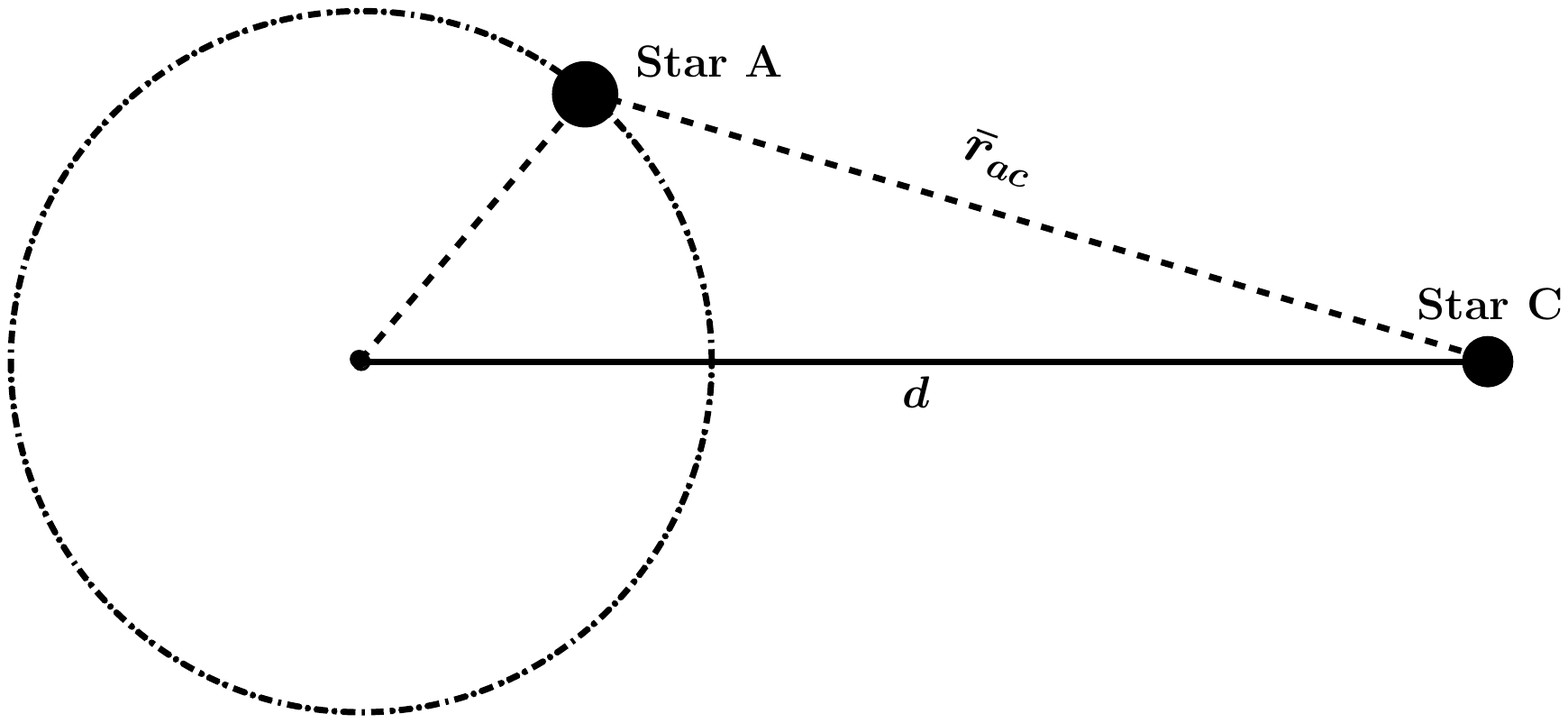}
	\end{figure}

	\noindent From Eq. \ref{eq:position ab} and Eq. \ref{eq:position bc} the position vector of star C observed from the star A (Fig. \ref{fig:Aperant-motion}) is expressed as
	
	\[
	\vec{r}_{ac}=\vec{r}_{ab}+\vec{r}_{bc}=-(R\sin\omega_{1}t+r\sin\omega t)\:\hat{i}-(R\cos\omega_{1}t+r\cos\omega t)\:\hat{j}.
	\]
	
	\noindent Therefore, the position vector of star A observed from the star C (Fig. \ref{fig:Aperant-motion}) is obtained as
	
	\[
	\vec{r}_{ca}=-\vec{r}_{ac}=(R\sin\omega_{1}t+r\sin\omega t)\:\hat{i}+(R\cos\omega_{1}t+r\cos\omega t)\:\hat{j.}
	\]
	
	\section{\label{gwmergerke}Gravitational waves from the system and loss of kinetic energy}
	
	The gravitational wave is generated as a result of time varying curvature in space-time \cite{quadrupole,Theory_of_Gravitational_Waves,The_basics_of_gravitational_wave_theory}. The resulting disturbance propagates as a ripple in the fabric of space-time with the speed of light. Since mass and momentum are conserved here, the monopole and the dipole modes of the radiation are absent. The wave is quadrupolar in nature and has two polarization states denoted as \textquotedblleft $+$\textquotedblright{} and \textquotedblleft $\times$\textquotedblright{} polarization. According to Einstein's quadrupole formula for gravitational radiation \cite{wheeler,schutz,h_and_L,tapir_caltech} the wave amplitude at a field point is inversely proportional to the distance from the source and directly proportional to the second time derivative of the quadrupole moment of the source.
	
	Assuming gravitational waves are weak, the metric can be expressed as,
	\[
	g^{ij}=\eta^{ij}+h^{ij},
	\]
	
	\noindent where $\eta^{ij}$ is the metric of Minkowski space-time and $h^{ij}$ is the metric perturbation tensor, which describes the gravitational wave. The $(ij)^{th}$ component of perturbation $h$ is given by \cite{wheeler,schutz,h_and_L,tapir_caltech},
	\begin{equation}
	h^{ij}=\frac{2G}{d\,c^{4}}\frac{\mathrm{d}^{2}}{\mathrm{d}t^{2}}[Q_{ij}],\label{eq:h}
	\end{equation}
	where $Q_{ij}$ is the quadrupole moment of the system. In order to estimate gravitational wave luminosity experienced by star B or star C due to their own motion in presence of star A, the apparent quadrupole moment ($Q$) of the star A relative to star B or star C has to be evaluated first. The $(ij)^{\text{th}}$ components of the quadrupole moment tensor $Q_{ij}$ is given by,
	
	\[
		Q_{ij}=-(I_{ij}-\frac{1}{3}\delta_{ij}Tr\:I),\label{eq:quadrupole1}
	\]
	
	\noindent where $Tr(I)$ is $\frac{6}{5}MR^{2}+2Mr^{2}$ and $I_{ij}$ is the $(ij)^{th}$ component of the moment of inertia tensor. The possible components of quadrupole moment tensor are given below in Eq. \ref{eq:qij} is tabulated in the following table.
	
	\begin{eqnarray}
		\begin{array}{ccccccc}
		Q_{XX} & = & Mr^{2}(\frac{2}{3}-\cos^{2}\omega t),&   &
		Q_{YY} & = & Mr^{2}(\frac{2}{3}-\sin^{2}\omega t),\\
		Q_{ZZ} & = & -\frac{1}{3}Mr^{2},&   &
		Q_{XY} & = & -\frac{1}{2}Mr^{2}\sin2\omega t,\\
		Q_{YX} & = & -\frac{1}{2}Mr^{2}\sin2\omega t,&   &
		Q_{YZ} & = & 0,\\
		Q_{ZY} & = & 0,&   &
		Q_{ZX} & = & 0,\\
		Q_{XZ} & = & 0.
		\end{array}
		\label{eq:qij}
	\end{eqnarray}
	
	\noindent Substituting the above expressions in (Eq. \ref{eq:quadrupole1}) we get the quadrupole moment tensor $Q$ as
	
	\begin{equation}
	Q=Mr^{2}\left[\begin{array}{ccc}
	\frac{2}{3}-\cos^{2}\omega t & -\frac{1}{2}\sin2\omega t & 0\\
	-\frac{1}{2}\sin2\omega t & \frac{2}{3}-\sin^{2}\omega t & 0\\
	0 & 0 & -\frac{1}{3}
	\end{array}\right].\label{quadrupole}
	\end{equation}
	
	\noindent Using Eq. \ref{quadrupole} and Eq. \ref{eq:h}, the perturbation matrix $h$ (with component $h^{ij}$)
	
	\[
	h=\frac{4GMr^{2}\omega^{2}}{dc^{4}}\left[\begin{array}{ccc}
	\cos2\omega(t-\frac{d}{c}) & \sin2\omega(t-\frac{d}{c}) & 0\\
	\sin2\omega(t-\frac{d}{c}) & -\cos2\omega(t-\frac{d}{c}) & 0\\
	0 & 0 & 0
	\end{array}\right].
	\]
	
	\noindent Differentiating Eq. \ref{quadrupole} thrice with respect to $t$, we obtain
	
	\begin{equation}
	\dddot{Q}=4Mr^{2}\omega^{3}\left[\begin{array}{ccc}
	-\sin2\omega(t-\frac{d}{c}) & \cos2\omega(t-\frac{d}{c}) & 0\\
	\cos2\omega(t-\frac{d}{c}) & \sin2\omega(t-\frac{d}{c}) & 0\\
	0 & 0 & 0
	\end{array}\right].\label{A}
	\end{equation}
	
	\noindent The expression of luminosity is given by \cite{wheeler,schutz,h_and_L,tapir_caltech}
	
	\begin{equation}
	\mathcal{L}_{\text{GW}}=\frac{G}{5c^{5}}\left\langle\stackrel[k,n=1]{3}{\sum}\left(\dddot{Q}_{kn}\left(t-\frac{d}{c}\right)\dddot{Q}_{kn}\left(t-\frac{d}{c}\right)\right)\right\rangle,\label{eq:B}
	\end{equation}
	
	\noindent Substituting $\dddot{Q}$ from Eq. \ref{A} in Eq. \ref{eq:B} the expression for luminosity for the present case takes the form
	
	\begin{equation}
	\mathcal{L}_{\text{GW}}=\frac{32GM^{2}\omega^{6}r^{4}}{5c^{5}}.\label{eq:luminosity}
	\end{equation}
	
	Following our assumption $M\gg m$, the star A (massive star) can be considered as stationary and the whole binary system revolves around star A in a circular orbit. But both the stars B and C consisting of the binary system orbit each other (Fig. \ref{fig:Triple-star-system}). Therefore, an observer positioned on star B or star C will experience gravitational waves as a consequence of apparent motion of the star A. These gravitational waves will carry energy, which in this case is equal to the change of kinetic energy of both the constituents of the binary system  \cite{An_analysis_of_the_LIGO_discovery_based_on_Introductory_Physics,Are_gravitational_waves_spinning_down_PSRJ1023+0038,The_basic_physics_of_the_binary_black_hole_merger_GW150914}. As the kinetic energy decreases the mutual separation between the two stars of the binary system suffers depletion whereas the angular frequency of the binary system increases in order to conserve the angular momentum much the same way as the conventional BH-BH merging processes \cite{wheeler,schutz,h_and_L,tapir_caltech}. At any given time $t$ the kinetic energy ($E_{k}$) of star B or star C is given by,
	\begin{equation}
	E_{k}=\frac{1}{2}mc^{2}(\gamma-1)=\frac{\frac{1}{2}mc^{2}}{\sqrt{1-\frac{\omega^{2}r^{2}}{c^{2}}}},\label{eq:kinetic_energy}
	\end{equation}
	where, $m$ is the mass of star B or star C, $\omega$ is its angular velocity and $v$ is the linear velocity of star B or star C.
	
	Equating the rate of change of kinetic energy with time ($\dot{E}_{k}$) to the energy received per unit time in the form of gravitational wave, that can be observed by an observer at one of the binary stars, the following conservation equation is obtained as,
	
	\begin{equation}
	\mathcal{L}_{\text{GW}}\times\left(\frac{\pi\Re^{2}}{4\pi d^{2}}\right)+\frac{d}{dt}E_{k}=0,\label{eq:Conservation equation}
	\end{equation}
	
	\noindent where $\Re$ denotes the radius of both the stars forming binary system. Substituting for $E_{k}$ and $\mathcal{L}_{\text{GW}}$ from (Eq. \ref{eq:kinetic_energy}) and (Eq. \ref{eq:luminosity}) in (Eq. \ref{eq:Conservation equation}) one obtains the time required for the collapse.
	
	The above mentioned merger event is faster relative to a merger a similar binary system but without the  third massive star considered here. This is evident in the frame of either of the rotating binary stars (of equal mass). In this frame the distance to the central star changes and this is reflected in the change of the quadrupole moment of the system resulting in loss of energy through gravitational waves. However, a faraway inertial observer would find that the central star is almost unmoving, and there is no significant change in the quadrupole moment. Although this observer would not observe any gravitational waves, s/he would nonetheless observe the same speeding up of the merger due to the processes discussed above. This is because of the fact that as the binary stars orbit around their common center of mass in the presence of another massive object, both of the binary companion experience different amount of space-time curvature in a periodic fashion and the kinetic energy lost in the process is quantitatively same as the energy lost by the ``fictitious'' gravitational wave emission noted in the first instance. This process is analogous to the fictitious force \cite{fictitious} which appears to a classical non-inertial observer.
	
	\section{\label{csandrs}Calculations and results}
	
	We numerically solve Eq. \ref{eq:Conservation equation} to obtain our results. The representative values chosen for the present work are tabulated in Table \ref{table1}.
	
	\begin{figure}[H]
		\centering{}
		\caption{\label{fig:r vs t graph}The variation of mutual separation of star A and binary system with time for the adopted set of values given in Table \ref{table1}.}
		\includegraphics[width=10cm,height=6cm]{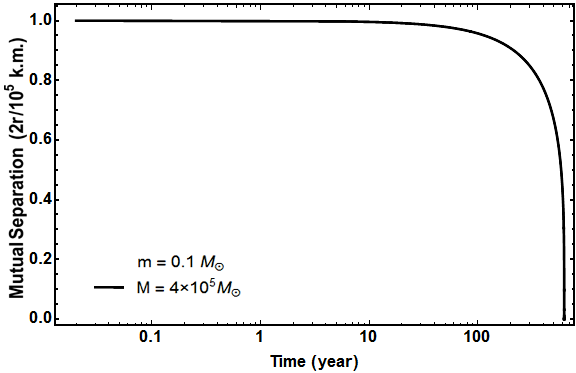}
	\end{figure}
	
\begin{figure}[H]
		\centering{}
		\caption{\label{fig:Comparison-between-two}Comparison between the mutual separation of the binary with the presence of a third massive body and without it.}
		\includegraphics[width=10cm,height=6cm]{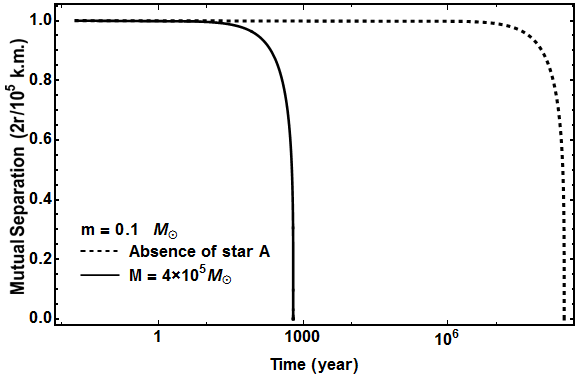}
	\end{figure}

	We first calculate the variation of mutual separation between the binary system and the central massive star with time to demonstrate how rapidly the binary system approaches the merger. This is shown in Fig. \ref{fig:r vs t graph}. From Fig. \ref{fig:r vs t graph} it is evident that for the chosen set of representative values (Table \ref{table1}), the mutual separation between the two stars of binary system remains constant initially and then suffers a rapid decrease. For the present case the mutual separation decreases sharply after $\sim50$ years. In Fig. \ref{fig:Comparison-between-two} we show similar variation (mutual separation with time) when the star A (Fig. \ref{fig:Triple-star-system}) is absent. We find that for the case of triple star system (star A, star B and star C) binary system of star B and star C merges after about $624.34$ years whereas, in the absence of star A the merger of the same binary system occurs after around $2.53\times10^{8}$ years.
	
	\begin{table}[H]
		\centering{}
		\caption{\label{table1}The various representative values chosen for the present calculations.}
		\begin{tabular}{ccccc}
			\hline
			\hline 
			\multicolumn{3}{c}{Mass (in $M_{\odot}$)} & \multirow{2}{*}{$r$ (in meter)} & \multirow{2}{*}{$d$ (in meter)}\tabularnewline
			\cline{1-3} 
			Star A & Star B & Star C &  & \tabularnewline
			\hline 
			$4\times10^{5}$ & $0.1$ & $0.1$ & $10^{8}$ & $10^{10}$\tabularnewline
			\hline
			\hline
		\end{tabular}
	\end{table}

\begin{figure}[H]
		\centering{}
		\caption{\label{fig:joint} The variation of frequency of the gravitational wave as seen from the binaries (B-C) during the final second of the merger, where the amplitude of the wave is indicated in the sidebar (a). Corresponding waveform at the final second before the merger (b). See text for details.}
		\begin{tabular}{cc}
			\includegraphics[width=8cm,height=6cm]{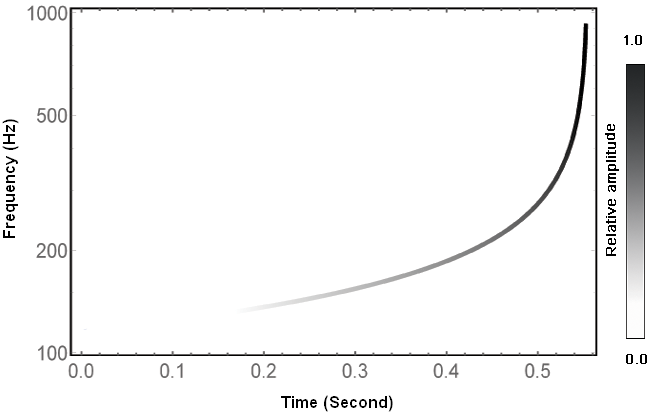} & \includegraphics[width=8cm,height=6cm]{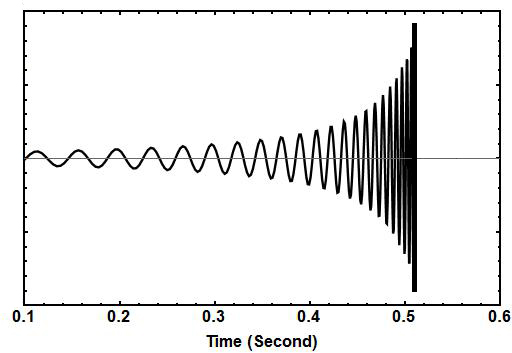}\tabularnewline
			(a) & (b)\tabularnewline
		\end{tabular}
	\end{figure}

	Due to the emission of gravitational waves, the stars belonging to binary would lose energy and continually spiral in. As a result the angular frequency of rotation of the stars in the system increases and eventually collapse and merge together. In Fig. \ref{fig:joint}a we plot the variation of this frequency with time during the last one second of the merger event. One sees that at a time of about $\sim0.15$ second before the merger the frequency starts increasing very rapidly. Fig. \ref{fig:joint}b also shows the relative amplitude of the gravitational waves as the binary system approaches the collapse. It is seen that the relative amplitude of the gravitational waves increases and tends to attain a maximum near the time of merger. The corresponding waveform is shown in Fig. \ref{fig:joint}b. Fig. \ref{fig:joint}b also shows how rapidly the angular frequency as well as the amplitude of the gravitational waves increase as the system approaches merger.
	
\begin{center}
		\begin{figure}[H]
		\centering{}
		\caption{\label{dependence}The mutual separation v/s time graph for different masses of star A (a) and that for different masses of binary stars (b). The merging time for different masses of star A (c) and the same for the binary stars (d).}
		\begin{tabular}{cc}
			\includegraphics[width=8cm,height=4cm]{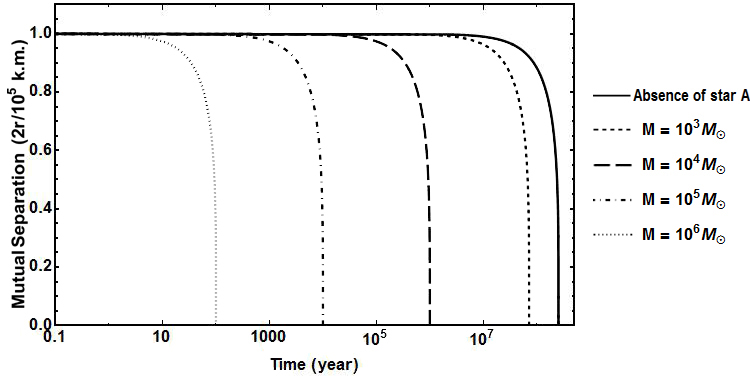} & \includegraphics[width=8cm,height=4cm]{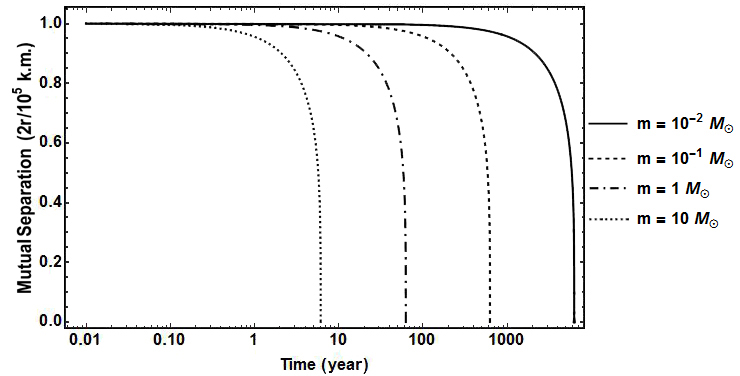}\tabularnewline
			(a) & (b)\tabularnewline
			& \tabularnewline
			\includegraphics[width=8cm,height=5cm]{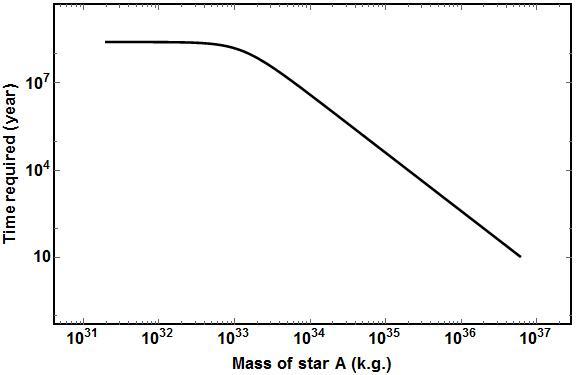} & \includegraphics[width=8cm,height=5cm]{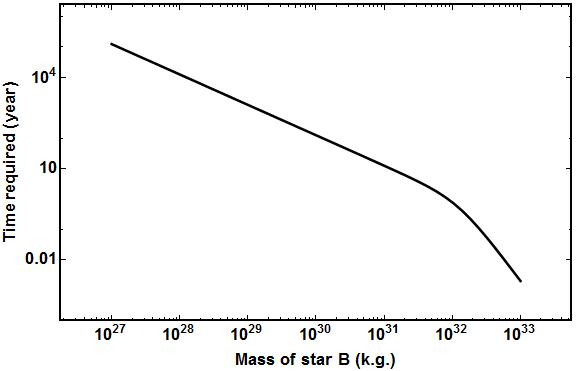}\tabularnewline
			(c) & (d)\tabularnewline
		\end{tabular}
		\end{figure}
	\end{center}

	Since the apparent luminosity ($\mathcal{L}_{\text{GW}}$) of the gravitational waves is a function of the masses ($m$) of the binary stars and the mass $M$ of the primary star, merging time ($t$) of the binary system varies significantly with them. The estimated time ($t$) required for the collision of the binary stars are tabulated in Table \ref{table2} for several representative combinations of the masses $m$ and $M$. The initial separation between the binary stars is fixed at a value of $2\times10^{8}\:\mathrm{m}$ for each case considering in Table \ref{table2}. The dependence of the merging time ($t$) on the masses $m$ and $M$ are described in Fig \ref{dependence}. In Fig. \ref{dependence}a we show the variations of the mutual separations between the binary members for different chosen masses of star A whereas Fig. \ref{dependence}b illustrates similar variations for different chosen masses of star B or star C.
	
	The effect of mass $M$ on the merging time of the binary is shown in Fig. \ref{dependence}c for a fixed value of $m=0.1\, M_{\odot}$. From Fig. \ref{dependence}c it is observed that as $M$ increases the merging time $t$ gradually decreases. It can also be seen from Fig. \ref{dependence}c that, initially the variation of $t$ is negligibly small for $M\lesssim 1000\, M_{\odot}$ and beyond this value of $M$ the merging time $t$ sharply decreases with the increase of $M$. Fig. \ref{dependence}d demonstrates similar variations of merging time with $m$ when the mass $M$ is fixed at a chosen value of $4\times 10^{5}\,M_{\odot}$. Fig. \ref{dependence}d also exhibits that the merging time consistently decreases with the increase of mass $m$ of one of the members of the binary system considered here, until upto $m\sim 50 M_{\odot}$ beyond which the merging time decreases more rapidly.
	\begin{center}
		\begin{table}[H]
			\centering{}
			\caption{\label{table2}}
			\begin{tabular}{cccc}
				\hline
				\hline 
				\textbf{\textit{\footnotesize{}$\begin{array}{c}
						\mathrm{Mass}\:\mathrm{of}\:\mathrm{each}\\
						\mathrm{binary}\:\mathrm{star}\\
						\mathrm{in}\:M_{\odot}
						\end{array}$}} & \textbf{\textit{\footnotesize{}$\begin{array}{c}
						\mathrm{Mass}\:\mathrm{of}\\
						\mathrm{central}\:\mathrm{BH}\\
						\mathrm{in}\:M_{\odot}
						\end{array}$}} & \textbf{\textit{\footnotesize{}$\begin{array}{c}
						\mathrm{Time}\:\mathrm{calculated}\\
						\mathrm{in}\:\mathrm{new}\:\mathrm{method}\\
						(\mathrm{in}\:\mathrm{year})
						\end{array}$}} & \textbf{\textit{\footnotesize{}$\begin{array}{c}
						\mathrm{Time}\:\mathrm{calculated}\\
						\mathrm{in}\:\mathrm{conventional}\\
						\mathrm{method}\:(\mathrm{in}\:\mathrm{years})
						\end{array}$}}\tabularnewline
				\hline 
				$0.05$ & $1\times10^{5}$ & $19978.60$ & $2.02733\times10^{9}$\tabularnewline 
				$0.10$ & $1\times10^{5}$ & $9988.99$ & $2.53416\times10^{8}$\tabularnewline
				$0.15$ & $1\times10^{5}$ & $6659.00$ & $7.50862\times10^{7}$\tabularnewline
				$0.05$ & $5\times10^{5}$ & $799.150$ & $2.02733\times10^{9}$\tabularnewline
				$0.10$ & $5\times10^{5}$ & $399.575$ & $2.53416\times10^{8}$\tabularnewline
				$0.15$ & $5\times10^{5}$ & $266.383$ & $7.50862\times10^{7}$\tabularnewline
				$0.05$ & $1\times10^{6}$ & $199.788$ & $2.02733\times10^{9}$\tabularnewline 
				$0.10$ & $1\times10^{6}$ & $99.8938$ & $2.53416\times10^{8}$\tabularnewline 
				$0.15$ & $1\times10^{6}$ & $66.5958$ & $7.50862\times10^{7}$\tabularnewline
				\hline
				\hline 
			\end{tabular}
		\end{table}
		\par\end{center}
	
	\section{\label{conclu}Conclusion}
	
	In this work we have studied gravitational wave emission for an effective three body system consisting a binary star and a remote black hole of intermediate black hole. As the stars in the binary revolve around each other, the distance between each of the stars and the black hole also undergoes a variation. As a consequence, there will be an apparent variation of the position of the black hole with respect to a frame fixed to any of the stars of the double star system. This induces a variation of space-time curvature resulting in the propagation of gravitational waves which could be experienced by an observer, who is at rest with respect to any of the stars of the system. The energy involved due to this apparent emission of the gravitational waves experienced by the fixed frame observer is compensated by the loss of kinetic energy of the binary system. As a result the stars of the binary will inspiral towards an eventual collapse. In this work the evolution of angular frequency and amplitude of such gravitational waves have been studied when the system approaches the merger.
	
	The contribution of special relativistic effect to the merging process is found to be of minimal relevance in the current work. We have also neglected the slight deformation of orbital motion of stars in binary systems as well as any effect resulting from shape deformations of the stars. A significant inference that has emerged from this work is that the time-scale of merging of a coalescing binary due to the emission of gravitational waves depend not only on the masses of the stars belonging to the system, but also on the mass of a neighboring star or a black-hole. Moreover, the above mentioned phenomena can also serve as a marker for detecting and identifying IMBH objects if they really exist. The mass of such an IMBH neighbor can be estimated from a measurement of the change of frequency of the binary system as well as its orbital radius with time. In this calculation we have also disregarded the spin motion of each component star presuming that they are tidally locked \cite{schutz,tidal}.
	
	In summary, during the course of investigating the effect of a `partner' on the emission of gravitational waves in a binary system, we have found a `speed up', a result which would have important consequences for observers interpreting gravitational wave signatures from deep space. 
	
	\section*{Acknowledgement}
	
	Two of the authors (S.B. and A.H.) wish to acknowledge the support received from St. Xavier's College Central Research Facility. A.H. also acknowledges the University Grant Commission 
	(UGC) of the Government of India, for providing financial
	support, in the form of UGC-CSIR NET-JRF.

\end{document}